\documentclass[10pt,a4paper]{article}
\usepackage{axodraw}

\begin{document}
\def\bqa{\begin{eqnarray}}
\def\eqa{\end{eqnarray}}
\def\bq{\begin{equation}}
\def\eq{\end{equation}}
\def\nll{\nonumber \\}
\newcommand{\sss}[1]{\scriptscriptstyle{#1}}
\newcommand{\ds }{\displaystyle}
\def\GF {G_{\sss F}}
\def\ml {m_{\ell}}
\def\mw {M_{\sss W}}
\def\gw {\Gamma_{\sss W}}
\def\mz {M_{\sss Z}}
\def\gz {\Gamma_{\sss Z}}
\def\mh {M_{\sss H}}
\def\stw{s_{\sss W}}
\def\ctw{c_{\sss W}}
\def\MSbar{\overline{\mathrm{MS}}}
\def\GeV{\unskip\,\mathrm{GeV}}
\def\MeV{\unskip\,\mathrm{MeV}}
\def\hs{\hat s}
\def\hspr{\hat{s}'}
\def\qmo{Q_\ell}
\def\mmo{m_\ell}
\def\qup{Q_{u}}
\def\qqu{Q_{q}}
\def\mup{m_{u}}
\def\mtp{m_{t}}
\def\mtpt{\tilde{m}_t}
\def\mbt{m_{b}}
\def\cmi{c_-}
\def\cpl{c_+}
\def\mqu{m_{q}}
\def\chic{\chi^*}
\def\order#1{{\mathcal O}\left(#1\right)}
\def\msbar{\overline{\tiny \mathrm{MS}}}
\def\Litwo{\mbox{${\rm{Li}}_{2}$}}
\def\als{\alpha_{_S}}
\def\alsS{\alpha^2_{_S}}
\def\wtp{\Gamma_t}
\def\mg{m_g}
\def\mwt {\widetilde{M}^2_{\sss{W}}}
\def\thle{\vartheta_{l}}
\def\cA{{\cal A}}
\def\SANC{{\tt SANC}~}
\def\mq{m_q}
\def\Qs{Q^2}
\def\mtpt{\tilde{m}_t}
\def\ieps{i\varepsilon}
\def\gdp{\gamma_{+}}
\def\gdm{\gamma_{-}}
\def\gdpm{\gamma_{\pm}}

% from Brem
\def\spr{s'}
\def\sqrspr{\sqrt{s'}}
\def\gtp{\gamma_t}
\newcommand{\ImJbW}{ImJbW}
\newcommand{\ReJbW}{ReJbW}
\newcommand{\koef}{k_0} 
\newcommand{\koeff}{K}
\newcommand{\kpl}{k^+}
\newcommand{\kmi}{k^-}
\def\rst{r_{st}}
\def\ome{\bar{\omega}}

\begin{center}
%{\bfseries On the NLO QCD Radiative Corrections to Single-top Production}
{\bfseries Standard SANC modules for NLO QCD Radiative Corrections to Single-top Production}
\footnote{\small
This work is partly supported by RFFI grant $N^{o}$10-02-01030-a}

\vskip 5mm

D. Bardin$^{*}$,
S. Bondarenko$^{*}$,
P. Christova$^{*}$,
L. Kalinovskaya$^{*,**}$,
V. Kolesnikov$^{*}$,\\
W. von Schlippe$^{\ddag}$,
K. Yordanova$^{\dag}$,

\vskip 5mm

{\small {\it $^{*}$ Joint Institute for
Nuclear Research, 141980 Dubna, Russia}} \\
{\small {\it $^{*,*}$ Higher Mathematics Department, The ``Dubna'' international University for
         Nature, Society, and Man, Dubna, 141980, Russia}}  \\
{\small {\it $^{\ddag}$ PNPI, 188300 St. Petersburg, Russia}}\\
{\small {\it $^{\dag}$Bishop Konstantin Preslavsky University, Shoumen, Bulgaria }}

\end{center}

\vskip 5mm

\centerline{\bfseries Abstract}
It this paper we present the results obtained with the newly created Standard \SANC~ \linebreak
modules
for calculation of the NLO QCD corrections to single top production processes in $s$
and $t$ channels at the partonic level, as well as top-decays. The main aim of these results
is to prove the correct work of modules. 
A comprehensive comparison with results of the {\tt CompHEP} system is given, where possible.
These modules are intended to be used in Monte Carlo 
generators for single top production processes at the LHC.
As in our recent paper, devoted to the electroweak corrections to these processes, we study 
the regularization of the top-legs associated infrared divergences with aid of the complex mass 
of the top quark. A comparison of QCD corrections with those computed by the conventional
method is presented both for top production and decays. For $s$ channel production we give
an analytic proof of equivalence of the two methods in the limit of low top width.


\vskip 10mm

\section{Introduction}
%---------------------
There is continued interest in precision calculations of single top quark
production cross sections for the Tevatron and LHC
(see {\it e.g.}~\cite{Bernreuther:2008ju} and references therein).
This is motivated to a large extent by the fact that this process is the only way of measuring 
the CKM matrix element $|V_{tb}|$ directly, providing a sensitive test of the
3-generation scheme of the Standard Model, and that it is an important
background to various processes beyond the SM, including Higgs boson production.

Moreover, such events are already observed 
at the Tevatron experiments D0~\cite{Abazov:2006gd,Abazov:2009ii} and
CDF~\cite{Aaltonen:2008sy,Aaltonen:2009jj}
and at the LHC~\cite{Chatrchyan:2011vp}.
There is a need to prepare software for  precision analysis of the high statistics
samples of single top quark events in future measurements at the LHC, running at 7 and 
14 TeV.

Most of the theoretical work on single top production has been concerned with
NLO QCD corrections (see {\it e.g.}~\cite{Harris:2002md}, \cite{Campbell:2009ss}), 
leading to the development of Monte Carlo generators, such as ZTOP~\cite{Sullivan:2004ie},
MC@NLO~\cite{Frixione:2005vw} or SingleTop~\cite{Boos:2006af},
incorporated in the standard LHC tools.

In our previous paper \cite{Bardin:2010mz} we have presented the calculation of 
the cross sections with one-loop electroweak corrections, regulating the infrared divergences (IRD)
associated with the top-quark leg by introducing the width of the top quark.

 In the present paper we extend that approach to the calculation of
one-loop QCD corrections in the spirit of the complex-mass scheme~\cite{Denner:1999gp}.
To our knowledge there is no other work utilizing this approach to the IRD regularization
among the many papers dealing with higher order QCD corrections to
single top quark production, see {\it e.g.} \cite{Kidonakis:2011wy} and
references therein, as well as papers devoted to the complex-mass scheme 
itself, e.g.~\cite{Denner:2006ic},~\cite{Actis:2008uh}.

This paper should be read together with \cite{Bardin:2010mz}, whose structure
we have taken over here.
 In section \ref{Amplitudes} %%section 2:
we recall the covariant amplitude for all processes $t\bar{b}\bar{u}d\to0$
and $\bar{t}{b}{u}\bar{d}\to0$ and show the conversions to the processes of
$t$ and $\bar{t}$ decay and to $s$  and $t$ channel single top quark
production, and point out the essential differences between the EW and QCD
formulations of these processes.
 In section \ref{BVudtb}     %% section 3:
we discuss in detail the regularization of single $t$ and $\bar{t}$ quark
production in the conventional approach --- with zero top quark width --- and
in the complex top quark mass approach. We present explicit formulas derived
for both cases.
In section \ref{ResAndComp}  %% section 4:
we present the results of numerical calculations which show the stability
against variations of the soft-hard separation parameter and of the top quark
width.
Finally, in section 5 we present our conclusions and outlook to further work
on single top quark production within the framework of SANC,
aimed at the creation of an MC generator that  simultaneously takes account 
of NLO EW and QCD corrections in the so-called 5F-scheme~\cite{Campbell:2009ss}.

\clearpage

\section{Covariant Amplitude \label{Amplitudes}}
%-----------------------------------------------
In this section we proceed in the spirit of the ``multi-channel approach''
developed in  % see
Refs.~\cite{Andonov:2004hi} and~\cite{Bardin:2007wb}, and follow closely % to
the presentation of our previous paper~\cite{Bardin:2010mz}.
First, we consider annihilation into vacuum with all particles incoming.

\subsection{All particles incoming}
%----------------------------------
In QCD, there is no contribution from box diagrams such as those
of~\cite{Bardin:2010mz} Figs.~1,2 which were appropriate for the EW case. Therefore
in the present case we have a sum of two one-loop vertices
instead of the two vacuum diagrams of~\cite{Bardin:2010mz}:
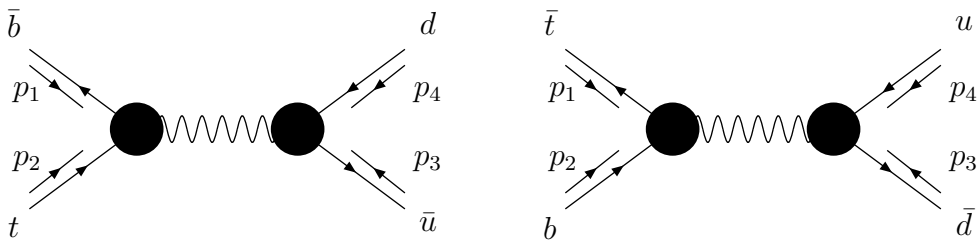
\begin{figure}[!ht]
\begin{center}
\begin{tabular}{cc}
   \begin{picture}(125,80)(210,0)
%    \GOval(270,40)(34,5)(90){0.02}
     \ArrowLine(240,40)(200,70)
     \ArrowLine(200,10)(240,40)
     \ArrowLine(340,70)(300,40)
     \ArrowLine(300,40)(340,10)
     \Vertex(240,40){10}
     \Vertex(300,40){10}
     \Photon(240,40)(300,40){5}{8}

     \ArrowLine(200,16)(220,32)
     \ArrowLine(200,64)(220,48)
     \ArrowLine(340,16)(320,32)
     \ArrowLine(340,64)(320,48)

     \Text(200,54)[]{\large $p_1$}
     \Text(200,28)[]{\large $p_2$}
     \Text(350,28)[]{\large $p_3$}
     \Text(350,54)[]{\large $p_4$}

     \Text(195,80)[]{\large$\bar{b}$}
     \Text(195,3)[]{\large $t$}
     \Text(350,5)[]{\large $\bar u$}
     \Text(350,80)[]{\large$d$}
   \end{picture}
\hspace*{10mm}
&
\hspace*{10mm}
   \begin{picture}(125,80)(210,0)
%    \GOval(270,40)(34,5)(90){0.02}
     \ArrowLine(240,40)(200,70)
     \ArrowLine(200,10)(240,40)
     \ArrowLine(340,70)(300,40)
     \ArrowLine(300,40)(340,10)
     \Vertex(240,40){10}
     \Vertex(300,40){10}
     \Photon(240,40)(300,40){5}{8}

     \ArrowLine(200,16)(220,32)
     \ArrowLine(200,64)(220,48)
     \ArrowLine(340,16)(320,32)
     \ArrowLine(340,64)(320,48)

     \Text(200,54)[]{\large $p_1$}
     \Text(200,28)[]{\large $p_2$}
     \Text(350,28)[]{\large $p_3$}
     \Text(350,54)[]{\large $p_4$}

     \Text(195,80)[]{\large$\bar{t}$}
     \Text(195,3)[]{\large $b$}
     \Text(350,5)[]{\large $\bar d$}
     \Text(350,80)[]{\large$u$}
   \end{picture}
\end{tabular}
\end{center}
\caption[The $t \bar{b} \bar{u} d\to 0$ and $\bar{t} b  u \bar{d} \to 0$ processes]
        {The $t \bar{b} \bar{u} d\to 0$ and  $\bar{t} b  u \bar{d} \to 0$ processes}
\label{Diagram_tdec_vac}
\end{figure}

The Covariant Amplitudes (CA) of Figs.~\ref{Diagram_tdec_vac} are
characterized by four different structures and scalar Form Factors (FF)
if the masses of the light quarks but not the $b$ quark mass are neglected.
A common expression for this CA in terms of four scalar form factors, 
${\cal F}_{\sss{LL,RL,LD,RD}}(s,t)$, was presented in Ref.~\cite{Andonov:2004hi}.
We recall it here to introduce the notation:
\bqa
{\cal A} &=&i\,e^2\,\frac{d_{\sss W}(s)}{4}\Big[
  \gamma_\mu\gdp \otimes \gamma_\mu\gdp {\cal F}_{\sss{LL}}(s,t)
+ \gamma_\mu\gdm \otimes \gamma_\mu\gdp {\cal F}_{\sss{RL}}(s,t)\,
\nll
&&+\gdp \otimes \gamma_\mu\gdp(-i D_\mu){\cal F}_{\sss{LD}}(s,t)
+  \gdm \otimes \gamma_\mu\gdp(-i D_\mu){\cal F}_{\sss{RD}}(s,t)\Big],
\eqa
%---
where
\bq
\gdpm=\left( 1\pm \gamma_5 \right),\qquad
D_\mu=(p_1-p_2)_\mu\,.
\eq
%--
4-momentum conservation reads
\bq
p_1+p_2+p_3+p_4=0\,,
\eq
%--
and the invariants are defined by
\bq
s=-(p_1+p_2)^2,\quad t=-(p_2+p_3)^2,\quad u=-(p_1+p_3)^2,
\eq
%--
with
\bq
s+t+u=m^2_t+m^2_b\,.
\eq
%--
Furthermore,
\bq
d_{\sss W}(s)=\frac{V}{2\stw^2}\,\frac{1}{s-\mw^2+i\mw\Gamma_{\sss W}}\,,
\eq
%--
where $V=V_{tb}V_{ud}$ is the relevant product of CKM matrix elements.
The scalar form factors ${\cal{F}}$ are labeled according to
their structures, see~\cite{Andonov:2004hi}.

\subsection{Conversion to top decay}
%-----------------------------------
The CA for the decay $t(p_2) \to b(p_1) + u(p_3) + \bar{d}(p_4)$
is derived from the left Fig.~\ref{Diagram_tdec_vac} 
by the following 4-momentum replacement:
\[
  \begin{array}{llllllll}
    & p_1 & \to & -p_1,\quad&\quad  p_3 & \to & -p_3, \\[-1mm]
    & p_2 & \to &~~p_2,\quad&\quad  p_4 & \to & -p_4. \\
  \end{array}
\]

For the decay $\bar{t}(p_1) \to \bar{b}(p_2) + d(p_3) + \bar{u}(p_4)$ it
is more transparent to make the replacement from the right
Fig.~\ref{Diagram_tdec_vac}:
\[
  \begin{array}{llllllll}
    & p_1 & \to &~~p_1,\quad&\quad  p_3 & \to & -p_3, \\[-1mm]
    & p_2 & \to & -p_2,\quad&\quad  p_4 & \to & -p_4. \\
  \end{array}
\]
The corresponding two physical subprocess diagrams are those
shown in Fig.~3,4 of~\cite{Bardin:2010mz}. 

\subsection{Conversion to $s$ channel}
%-------------------------------------
The CA for the $s$ channel single-top production processes
$\bar{u}(p_1)+d(p_2)\to b(p_3)+\bar{t}(p_4)$ is obtained from the left
Fig.~\ref{Diagram_tdec_vac} by the replacement:
\[
  \begin{array}{llllllll}
    & p_1 & \to & -p_3,\quad&\quad  p_3 & \to &~~p_1, \\[-1mm]
    & p_2 & \to & -p_4,\quad&\quad  p_4 & \to &~~p_2, \\
  \end{array}
\]

\noindent
and for the processes $\bar{d}(p_1)+u(p_2)\to t(p_3)+\bar{b}(p_4)$ from
the right Fig.~\ref{Diagram_tdec_vac} by the conversion:
\[
  \begin{array}{llllllll}
    & p_1 & \to & -p_3,\quad&\quad  p_3 & \to &~~p_1, \\[-1mm]
    & p_2 & \to & -p_4,\quad&\quad  p_4 & \to &~~p_2. \\
  \end{array}
\]
As a result we get the two physical subprocess diagrams
shown in Fig.~5,6 of~\cite{Bardin:2010mz}. 

\subsection{Conversion to $t$ channel processes}
%-----------------------------------------------
In the CAs for the $t$ channel single-top production processes
$\bar{b}(p_1)+\bar{u}(p_2)\to\bar{d}(p_3)+\bar{t}(p_4)$ and
$\bar{b}(p_1)+d(p_2)\to u(p_3)+\bar{t}(p_4)$
it is convenient to make the replacement ``in pairs''.
From the left Fig.~\ref{Diagram_tdec_vac} one may perform two 4-momentum replacements:
\[
  \begin{array}{llllllll}
    &\hspace{7mm} p_1 & \to &~~p_1,\hspace{23mm} & p_1 & \to &~~p_1,\\[-1mm]
    &\hspace{7mm} p_2 & \to & -p_4,\hspace{23mm} & p_2 & \to & -p_4,\\[-1mm]
    &\hspace{7mm} p_3 & \to &~~p_2,\hspace{23mm} & p_3 & \to & -p_3,\\[-1mm]
    &\hspace{7mm} p_4 & \to & -p_3,\hspace{23mm} & p_4 & \to &~~p_2,\\
  \end{array}
\]

\noindent
which give rise to two different physical $t$ channel processes
as shown in Fig.~7 of~\cite{Bardin:2010mz}.

For the processes $b(p_1)+u(p_2)\to t(p_4)+d(p_3)$ and
 $b(p_1)+\bar{d}(p_2)\to t(p_4)+\bar{u}(p_3)$,
the pair of replacements from the right Fig.~\ref{Diagram_tdec_vac},
\[
  \begin{array}{llllllll}
    &\hspace{7mm} p_1 & \to & -p_4,\hspace{23mm} & p_1 & \to & -p_4,\\[-1mm]
    &\hspace{7mm} p_2 & \to &~~p_1,\hspace{23mm} & p_2 & \to &~~p_1,\\[-1mm]
    &\hspace{7mm} p_3 & \to & -p_3,\hspace{23mm} & p_3 & \to &~~p_2,\\[-1mm]
    &\hspace{7mm} p_4 & \to &~~p_2,\hspace{23mm} & p_4 & \to & -p_3,\\
  \end{array}
\]

\noindent
gives the corresponding pair of symbolic diagrams for the two remaining
$t$ channel processes. They are shown in Fig.~8 of~\cite{Bardin:2010mz}.

\clearpage

\section{Process $ud\to tb$ at NLO QCD\label{BVudtb}}
%----------------------------------------------------
\subsection{Notation and terminology}
%------------------------------------
After conversion of the vacuum annihilation diagrams of
Fig.~\ref{Diagram_tdec_vac} to the $s$ channel,
we will have two physical annihilation diagrams for both $t$ and $\bar{t}$ production (cf. symbolic
diagrams in Fig.~5--6 of Ref.~\cite{Bardin:2010mz}). For our purpose it is sufficient to consider
the top production NLO diagrams of Fig.~6 which, in turn, consist of ISR and FSR NLO contributions
like the two fat vertices in diagrams in Fig.~\ref{Diagram_tdec_vac}, standing for emission (and
reabsorption) of a real (or virtual) gluon (cf. diagrams in Fig.~9,10 and 12 of 
Ref.~\cite{Bardin:2010mz}).

\begin{figure}[!ht]
\begin{center}
   \begin{picture}(125,80)(210,0)
%    \GOval(270,40)(34,5)(90){0.02}
     \ArrowLine(240,40)(200,70)
     \ArrowLine(200,10)(240,40)
     \ArrowLine(340,70)(300,40)
     \ArrowLine(300,40)(340,10)
     \Vertex(240,40){1}
     \Vertex(300,40){10}
     \Photon(240,40)(300,40){5}{8}

     \ArrowLine(200,16)(220,32)
     \ArrowLine(200,64)(220,48)
     \ArrowLine(320,32)(340,16)
     \ArrowLine(320,48)(340,64)

     \Text(200,54)[]{\large $p_1$}
     \Text(200,28)[]{\large $p_2$}
     \Text(350,28)[]{\large $p_3$}
     \Text(350,54)[]{\large $p_4$}

     \Text(195,80)[]{\large$\bar{d}$}
     \Text(195,3)[]{\large $u$}
     \Text(350,5)[]{\large $t$}
     \Text(350,80)[]{\large$\bar b$}
   \end{picture}
\end{center}
\caption[The FSR in the $\bar{d} u\to t\bar{b}(g)$ processes]
        {The FSR in the $\bar{d} u\to t\bar{b}(g)$ processes}
\label{Diagram_schannel_pl}
\end{figure}
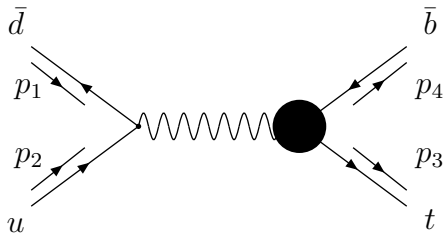

Since we are interested in the effects of the top quark width, the subject of our study
will be only the symbolic FSR NLO
diagram of Fig.~\ref{Diagram_schannel_pl}. In the following, the label FSR will be
assumed and omitted. It should be emphasized also that all the formulae of this Section refer to the
limit $\mbt\to 0$. 

We will use the following common terminology:\\ 
--- LO (or Born); \\
--- NLO=Virtual~+~Real;\\
--- \hspace{25mm}~Real=Soft~+~Hard;\\
--- SV~~=Soft~+~Virtual (with 2$\to$2 phase space);\\
--- NLO=SV+HA (Hard Bremsstrahlung with 2$\to$3 phase space).

\subsection{Conventional approach}
%---------------------------------
In the conventional approach (stable particles) one introduces a soft--hard separation parameter
\bq
{\ome}=1-\frac{s'_{\rm max}}{s}\,, \qquad
E^{\rm min}_g = \frac{\sqrt{s}}{2}{\ome}\,,
\eq
with invariants $s=M^2_{a,b}$ and $s'=M^2_{c,d}$ for the radiative processes $a+b\to c+d+g$.

And for the total NLO cross sections we have
\bq
\sigma^{\rm{NLO}}(s)=\sigma^{\rm{SV}}(s,\ome)+\sigma^{\rm{HA}}(s,\ome).
\label{sigmaNLO}
\eq
The infrared divergence cancels inside $\sigma^{\rm{SV}}=\sigma^{\rm{S+V}}$ no matter how it is
regularized (dimensionally or by a fictitious photon mass $\lambda$), while
the $\ome$ divergence 
cancels in the sum $\sigma^{\rm{NLO}}=\sigma^{\rm{SV+HA}}$, so that $\sigma^{\rm{NLO}}$ is finite.

At the end of this paragraph, we list the individual contributions.

\noindent
Soft+Virtual:
\begin{eqnarray}
\sigma^{\rm{SV}}(s,\ome)&=& \koef\Biggl\{ \sigma^{\rm Born}(s) \Biggl[-2
       + \left(L_s-2\right)\ln\left(\ome\right)
       + \left(\frac{3}{4}-L_{t1}\right)L_b
\nll &&
       + \left(\frac{3}{2}\frac{\mtp^2}{s_t}+\frac{5}{2}-2L_{t1}\right)L_{t1}
       + \left(\frac{5}{2}\frac{\mtp^2}{s_t}+\frac{3}{4}-2L_{t1}\right)L_t
       + 2\Litwo\left(\frac{\mtp^2}{s}\right)
\Biggr]
\nll &&
       + 2\Litwo\left(1\right) - \frac{9}{4}\koeff\mtp^2\frac{s_t}{s}\left(L_{t1}+L_t\right)
\Biggr\}.
\label{SVone}
\end{eqnarray}
Hard:
\begin{eqnarray}
\label{HAone}
\sigma^{\rm{HA}}(s,\ome)&=&\koef\Biggl\{\sigma^{\rm Born}(s)\Biggl[\frac{9}{4}-\frac{\mtp^2}{s_t}
       - \left(L_s-2\right)\ln\left(\ome\right)
       - \left(\frac{3}{4}-L_{t1}\right)L_b
\\ &&
 - \left(\frac{7}{2}-2L_{t1}\right)L_{t1}-\left(\frac{3}{4}-L_{t1}\right)L_t-2\Litwo\left(1\right) 
\Biggr]
       + \frac{1}{2}\koeff\left(s_t+2\mtp^2L_t\right)
\Biggr\}.
\nonumber
\end{eqnarray}
Born or LO :
\begin{eqnarray}
\sigma^{\rm Born}(s)&=&\frac{1}{2}\koeff\frac{s_t^2}{s^2}(2s+\mtp^2).
\end{eqnarray}

Here and below:
\begin{eqnarray}
&& 
\koef  = \frac{\alpha_s C_f}{\pi}\,, \qquad
\koeff = \frac{G_F^2\mw^4V^2}{6\pi\left[(s-\mw^2)^2+\mw^2\gw^2\right]}\,, 
\nll
&& L_s=2\ln\left(\frac{s_t}{\mtp\mbt}\right), \quad 
   L_t=\ln\left(\frac{s}{\mtp^2}\right),     \quad 
   L_b=\ln\left(\frac{s}{\mbt^2}\right),     \quad 
   L_{t1}=\ln\left(1-\frac{\mtp^2}{s}\right),\quad 
\nll
&& s_t=s-\mtp^2.
\end{eqnarray}

In the NLO cross section all divergences, including $b$ mass collinear divergences, cancel.
The final expression is finite and very compact:
\begin{eqnarray}
\sigma^{\rm{NLO}}(s)&=&\koef\Biggl\{\sigma^{\rm Born}(s)\Biggl[\frac{3}{4}
       -(L_t+1)L_{t1}+2\Litwo\left(\frac{\mtp^2}{s}\right) 
\Biggr]
\nll &&
       + \frac{1}{4}\koeff\frac{\mtp^2}{s^2}\Biggl[
         \left(5s_t\left(s+\mtp^2\right)+4\mtp^2s\right)L_t-3s_t^2\left(L_{t1}+1\right)-4\mtp^2s_t
\Biggr] 
\Biggr\}.\quad
\label{NLOone}
\end{eqnarray}

\subsection{Infrared regularization by the complex top quark mass} 
%-----------------------------------------------------------------
The main aim of this paragraph is to show that the infrared regularization by the complex top quark
mass leads exactly to the same final result, Eq.~(\ref{NLOone}), though the partial expressions for
SV and HA have a completely different form. For the introduction to the problem, we refer 
the reader to Section 3 of Ref.~\cite{Bardin:2010mz}.
Here we again only list the individual contributions to the total cross sections.\\
Soft+Virtual:
\begin{eqnarray}
\sigma^{\rm{SV}}(s,\ome,\wtp)&=& \koef\Biggl\{ \sigma^{\rm Born}(s)\Biggl[-1
   - \left(1-L_s\right)\ln\left(\frac{\wtp}{\mtp}\right)
   - \ln\left(\ome\right)
\nll &&
   + \left(\frac{3}{4}-L_{t1}-L_t\right)L_b
   + \left(\frac{3}{4}-3L_{t1}-\frac{1}{2}L_t\right)L_t+\left(\frac{5}{2}-2 L_{t1}\right)L_{t1}
\nll &&
   + \Litwo\left(\frac{\mtp^2}{s}\right)
   + 3\Litwo\left(1\right) \Biggr]
       - \frac{3}{4}\koeff\mtp^2\frac{s_t^2}{s^2}\left(L_{t1}+L_t\right)
\Biggr\}.
\label{SVtwo}
\end{eqnarray}
Hard:
\begin{eqnarray}
\sigma^{\rm{HA}}(s,\ome,\wtp)&=& \koef\Biggl\{ \sigma^{\rm Born}(s)
\Biggl[\frac{5}{4}-\frac{\mtp^2}{s_t}
   + \left(1-L_s\right)\ln\left(\frac{\wtp}{\mtp}\right)
   + \ln\left(\ome\right)
\nll &&
   - \left(\frac{3}{4}-L_{t1}-L_t\right)L_b
   - \left(\frac{7}{2}-2 L_{t1}\right) L_{t1}
\nll &&
   - \left(\frac{3}{4}-\frac{\mtp^2}{s_t}-2 L_{t1}-\frac{1}{2} L_t\right) L_t
   + \Litwo\left(\frac{\mtp^2}{s}\right)-3\Litwo\left(1\right) 
\Biggr]
\nll &&
   + \frac{1}{2}\koeff\left(s_t+2\mtp^2 L_t\right) 
\Biggr\}.
\label{HAtwo}
\end{eqnarray}

We note the presence of two regularization parameters: $\ome$ is due to conventional treatment of
infrared divergences associated with $b$ quark legs, and
$\ln\left({\wtp}/{\mtp}\right)$ is a new regularization parameter
associated with the $t$ quark leg, whose infrared divergence is regularized by its width. 
The individual (SV and HA contributions) have completely different form, but they sum up to exactly 
the same  finite NLO expression (\ref{NLOone}) in which all three types of divergent terms 
($\ln(\wtp/\mtp),\,\ln(\ome),\,\ln(\mbt)$) have canceled.

\subsection{Relevant AV functions}
%---------------------------------
The virtual contributions (due to vertex, Fig.~10, and the counter-term, Fig.~9, 
of Ref.~\cite{Bardin:2010mz}) were computed using the standard Passarino--Veltman
reduction~\cite{Passarino:1978jh}.
The relevant $C_0$ function 
\begin{equation}
C_0\left(-\mtp^2,\,-\mbt^2,\,Q^2;\,\mtpt,\,0,\,\mbt\right),
\end{equation}
with complex argument
\begin{equation}
 \mtpt^2=\mtp^2+\Delta_t\,,\quad\Delta_t\;=\;-i\mtp\wtp,
\end{equation}
is given by Eqs.~(22)--(24) of~\cite{Bardin:2010mz} and its limit 
at $\mbt \to 0$ by Eqs.~(25)--(26) of that paper.
Below we present its real part double limit: $\mbt \to 0$ and 
$\wtp \to 0$, which was used in the derivation of Eq.~(\ref{SVtwo}):
\begin{eqnarray}
&&\Re C_0\left(-\mtp^2,-\mbt^2,-s;\mtpt,0,\mbt\right)=\frac{1}{s_t}\Biggl\{
\Biggl[\ln\left(\frac{\wtp}{\mtp}\right)-L_{t1}-L_t
\Biggr]\left(L_b+L_t+2L_{t1}\right)\qquad
\\ &&\hspace*{50mm}
           +L_{t1} L_t+\frac{1}{2} L_t^2-\Litwo\left(1-\frac{\mtp^2}{s}\right)+4\Litwo(1)
\Biggr\},\quad
\nonumber
\end{eqnarray}
In the calculation of the counter-term, one meets the real part of the derivative of a $B_0$ function:
\begin{eqnarray}
\Re\left[B'_0\left(-\mtp^2,0,\mtpt\right)\right]=
\frac{1}{\mtp^2}\ln\left(\frac{\wtp}{\mtp}\right).
\end{eqnarray}

\subsection{Differential in $\spr$ hard gluon radiation}
%-------------------------------------------------------
The hard gluon contributions, Eq.~(\ref{HAone}) and Eq.~(\ref{HAtwo}), were derived by 4-fold
integration in the 2$\to$3 process phase space over three angular variables
and over the invariant 
$\spr$, which varies within the limits $\mtp^2\leq \spr \leq \spr_{\rm max}$. 
In this paragraph we present the $\spr$-integrands for the two approaches under consideration.

\subsubsection{Conventional approach}
%------------------------------------

In the conventional approach, the expression for $\sigma^{\rm HA}(s,\spr)$ is well known:
\begin{eqnarray}
\sigma^{\rm HA}(s,\spr) =  \koef     \Biggl\{ 
\sigma^{\rm Born}(s) (2 - L_{\spr}) \left( \frac{1}{s_t} - \frac{1}{s-\spr} \right)  
-\koeff\frac{s-\spr}{s}\left(\frac{\spr_t}{\spr} - k_1 L_{\spr} \right)
 \Biggr\},
\label{HAsprone}
\end{eqnarray}
where
\begin{equation}
k_1 = \frac{1}{2}\left( 1 + \frac{1}{2}\frac{\mtp^2}{s} \right), \qquad
L_{\spr} = 2\ln\left(\frac{\spr_t}{\mtp\mbt}\right),\qquad
\spr_t  = \spr-\mtp^2.
\end{equation}
The $\ome$ divergent term $1/(s-\spr)$ after integration over $\spr$ gives rise to 
$\ln\left(\ome\right)$ terms in Eq.~(\ref{HAone}) which cancel the corresponding terms in
Eq.~(\ref{SVone}).

\subsubsection{Infrared regularization by the complex top quark mass}
%--------------------------------------------------------------------

The expression for $\sigma^{\rm HA}(s,\spr)$ is more complicated in this case: 
\begin{eqnarray}
 \sigma^{\rm HA}(s,\spr)&=& \koef\Biggl\{ \sigma^{\rm Born}(s) \Biggl[
 \frac{1}{s_t}-\frac{1}{s-\spr}+\frac{\kpl}{2} L'_{\spr} - \frac{1}{s_t}L_{\spr} 
\nll &&
- \frac{1}{2} ( I_1(\spr)+I_2(\spr)) -  \frac{\mtp^2 }{s_t} I_3(\spr) \Biggr]
-\koeff\frac{s-\spr}{s}\left(\frac{\spr_t}{\spr} - k_1 L_{\spr} \right)
 \Biggr\}.
\label{HAsprtwo}
\end{eqnarray}
Here,
\begin{eqnarray}
L'_{\spr}&=& L_{\spr} - \ln\left(\frac{\spr}{\mbt^2}\right),
\nll
k^{\pm}&=& \frac{1}{s-\spr+i\gtp} \pm \frac{1}{s-\spr-i\gtp}\,,
\nll
\gtp  &=& \wtp\mtp\,.
\end{eqnarray}
%%%%%%%%%%%%%%%%%%%%%%%%%%%%%%%%%%%%%%%%%%%%%%%%%%%%

%id gtp=0;
%id [1/(s-spr+i*gtp)+1/(s-spr-i*gtp)]*[s-spr]=2; * as proportional to gtp.

The three non-trivial objects to be integrated are:
\begin{eqnarray}
I_1(\spr) &=&  k^{+}\Re(J_b^W(0,\mtp,\sqrspr,i{\gtp})), 
\nll 
I_2(\spr) &=&i k^{-}\Im(J_b^W(0,\mtp,\sqrspr,i\gtp)),
\nll
I_3(\spr) &=& \frac{1}{\gtp}\Im(J_b^W(0,\mtp,\sqrspr,i\gtp)),
\end{eqnarray}
%%%%%%%%%%%%%%%%%%%%%%%%%%%%%%%%%%%%%%%%%%%%%%%%%%%%%%%%%%%%%%%%%%%%%%%
where
\begin{eqnarray}
J_b^W(0,\mtp,\sqrspr,i\gtp)=
\ln\left(\frac{i\gtp\spr+(s-\spr)\mtp^2}{i\gtp\spr+(s-\spr)\spr}\right),
\end{eqnarray}
and ``0'' in the argument list stands for the limit $\mbt\to 0$.

Taking the integrals
\bq
I_i(s)=\int^{s}_{\mtp^2}I_i(\spr)\ d\spr,\quad i=1,2,3,
\eq
we get
\begin{eqnarray}
I_1(s) &=& \ln\left( \frac{s}{\mtp^2}\right)
\left[ \ln\left(\frac{\wtp^2}{\mtp^2}\right)-2 \ln\left(\frac{s_t}{\mtp^2}\right) \right]
\nll &&
-  \Litwo\left(\frac{-s_t}{\mtp^2}\right)
+ 2\Litwo\left(\frac{s_t}{s}\right)
-  \Litwo\left(1+\frac{s}{\mtp^2}\right)
            +\frac{3}{2}\Litwo(1),
\nll 
I_2(s) &=&    -\Litwo\left(\frac{-s_t}{\mtp^2}\right)
            +\Litwo\left(1+\frac{s}{\mtp^2}\right)
            -\frac{3}{2} \Litwo(1),
\nll
I_3(s) &=& \frac{1}{\mtp^2} \Biggl[-s_t \ln\left(\frac{\wtp}{\mtp}\right)
        + s_t\ln\left(\frac{-s_t}{\mtp^2}\right)-\mtp^2\ln\left(\frac{s}{\mtp^2}\right)\Biggr].
\label{threeint}
\end{eqnarray}
%%%%%%%%%%%%%%%%%%%%%%%%%%%%%%%%%%%%%%%%%%%%%%%%%%%%%%%%%%%%%%%%%%%%%%%
Substituting Eqs.~(\ref{threeint}) into Eq.~(\ref{HAsprtwo}) we arrive at Eq.~(\ref{HAtwo}). 

\clearpage

\section{Numerical Results\label{ResAndComp}}
%--------------------------------------------
In this section we present the \SANC results for the cross sections of the single
top quark production processes and for the top quark decays. The tree level
contributions, both Born and single real gluon  emission, are compared with
{\tt CompHEP} versions v.4.4.3~\cite{CompHEP:443} and v.4.5.1~\cite{CompHEP:451}.
All numerical results for this section were produced
with the standard \SANC INPUT working in the $\alpha(0)$ EW scheme.

The \SANC input parameters set:
\bq 
\begin{array}[b]{lcllcllcl}
G_F & = & 1.16637 \times 10^{-5} \GeV^{-2}, & & & \\
\alpha(0) &=& 1/137.035999, &\alpha_s& = & 0.107 \\
\mw & = & 80.403 \GeV, & \gw & = & 2.141  \GeV, \\
\mz & = & 91.1876\GeV, & \gz & = & 2.4952 \GeV, \\ 
\mh & = & 120\GeV,     & \wtp& = & 1.5517 \GeV, \\
m_u & = & 62\;\MeV,    & m_d & = & 83\;\MeV,  \\
m_c & = & 1.5\;\GeV,   & m_s & = & 215\;\MeV, \\
m_t & = & 174.2\;\GeV  & m_b & = & 4.7\;\GeV, \\
|V_{ud}| &=&1, & |V_{cs}| &=& 1, \\
|V_{us}| &=&0, & |V_{cd}| &=& 0, \\ 
|V_{tb}| &=&1. & & &
\end{array}
\label{SANCinput}
\eq

Numbers used for the comparison with {\tt CompHEP} for the decay channels were produced 
with CompHEP version 4.4.3 and with version v.4.5.1 for the production channels. 
For this comparison we use the standard {\tt CompHEP} setup and the $\alpha(0)$ EW
scheme, however, we used $\alpha_s = 0.12201\,(Q=\mz)$ and $m_u=m_d=66~\MeV$.
Moreover, in both codes we use the ``fixed$_{}$ $t$ width scheme''. 

The structure, notation and terminology used of this Section are very similar to those 
of the corresponding Section 4 of Ref.~\cite{Bardin:2010mz}. 
We refer the reader to the paragraphs that follow Eq.~(33) and
do not repeat here Eqs.~(34)--(35) and the surrounding text of the latter paper.

\subsection{Decay channels $t \to b + u + \bar{d}$}
%---------------------------------------
\subsubsection{SANC--CompHEP comparison}
%---------------------------------------

In Table~\ref{table_tbud_hard_QCD} we present results of the {\tt SANC-CompHEP} comparison of 
radiative 
decay width $\Gamma^{\rm hard}$ in $\MeV$ for two cuts on the gluon energy: $E_{g}\geq 5,10~\GeV$ 
(in the top rest frame) and for two options: $\Gamma_{t}=0(\neq 0)$ and six channels pid=19--24 for 
the case of \SANC (pid=19,21,23 for $\bar{t}$ decays, pid=20,22,24 for $t$ decays).

\begin{table}[!h]
\begin{center}
\begin{tabular}{|l|l|l|l|l|l|}
\hline
$E_{g}$, GeV &\multicolumn{2}{|c|}{5} & \multicolumn{2}{|c|}{10} \\ \hline
&$\Gamma_{t} = 0$&$\Gamma_{t}\neq 0$&$\Gamma_{t} = 0$&$\Gamma_{t} \neq 0$ 
                                                                 \\ \hline\hline
\multicolumn{5}{|c|}{$t \rightarrow b +  e^+  + \bar\nu_e +g $}  \\ \hline
{\tt CompHEP}  &  72.29(1)  &  71.80(2)  &  47.24(2)  &  47.13(2) \\ \hline   
\SANC, pid=19  &  72.28(1)  &  71.79(1)  &  47.20(1)  &  47.10(1) \\ \hline
\SANC, pid=20  &  72.28(1)  &  71.79(1)  &  47.20(1)  &  47.10(1) \\ \hline\hline
\multicolumn{5}{|c|}{$t \rightarrow b + \mu^+ + \bar\nu_\mu + g$} \\ \hline
{\tt CompHEP}  &  72.29(2)  &  71.80(2)  &  47.24(2)  &  47.13(2) \\ \hline
\SANC, pid=21  &  72.28(1)  &  71.79(1)  &  47.20(1)  &  47.10(1) \\ \hline
\SANC, pid=22  &  72.28(1)  &  71.79(1)  &  47.20(1)  &  47.10(1) \\ \hline\hline
\multicolumn{5}{|c|}{$t \rightarrow b + u + \bar{d} +g $}        \\ \hline
{\tt CompHEP}  &  1296.9(6) &  1295.6(5) &  820.3(3)  &  819.7(3) \\ \hline
\SANC, pid=23  &  1296.3(1) &  1294.9(1) &  819.8(1)  &  819.5(1) \\ \hline
\SANC, pid=24  &  1296.3(1) &  1294.9(1) &  819.8(1)  &  819.5(1) \\ \hline
\end{tabular}
\end{center}
\caption{{\tt SANC-CompHEP} comparison of decay width $\Gamma^{\rm hard}$ in $\MeV$ in the $\alpha$ 
scheme for two options: $\Gamma_{t}=0(\neq 0)$; and for six channels pid=19--24.
\label{table_tbud_hard_QCD}}
\vspace*{-3mm}
\end{table}

The main aim of Table~\ref{table_tbud_hard_QCD} is to demonstrate, for the  first time, that 
there is good agreement between \SANC and {\tt CompHEP} not only for $\Gamma_{t}=0$, but also for
the $\Gamma_{t}\neq 0$ options ($\Gamma_{t}\neq 0$ always refers to the value
in Eq.~(\ref{SANCinput})).

%------------------------------------------
\subsubsection{An internal SANC comparison}
%------------------------------------------

In \SANC there are two modules for the computation of hard gluon bremsstrahlung processes: fully
differential (5d), to be used in Monte Carlo simulations, and two dimensional (2d) in invariant 
variables $s=-(p_t-p_b)^2$ and $s'=-(p_u+p_d)^2$, being analytically integrated over all angular 
variables.
In order to check numerically the correctness of the analytic angular integration, we present
Table~\ref{table_tbud_hard_5_2_QCD} which shows the self-consistency of \SANC calculations of the
radiative contribution. Note that here $E_g$ is the cut in the $s'$ compound
rest frame: this is why 
the numbers in Tables~\ref{table_tbud_hard_QCD} and~\ref{table_tbud_hard_5_2_QCD} are different. 

\begin{table}[!h]
\begin{center}
\begin{tabular}{|l|l|l|l|l|l|}
\hline
$E_{g}$, GeV &\multicolumn{2}{|c|}{5} & \multicolumn{2}{|c|}{10}  \\ \hline
&$\Gamma_{t} = 0$&$\Gamma_{t}\neq 0$&$\Gamma_{t} = 0$&$\Gamma_{t} \neq 0$ 
                                                                  \\ \hline\hline
\multicolumn{5}{|c|}{$t \rightarrow b + \mu^+ + \bar\nu_\mu + g$} \\ \hline
{\tt SANC-5d}, ~pid=22 & 96.446(1) & 94.459(1) & 67.492(1) & 67.085(1)  \\ \hline
{\tt SANC-2s}',~pid=22 & 96.446(1) & 94.459(1) & 67.493(1) & 67.085(1)  \\ \hline
\multicolumn{5}{|c|}{$t \rightarrow b + u + \bar{d} +g $}         \\ \hline
{\tt SANC-5d}, ~pid=24 & 1245.6(1) & 1239.6(1) & 763.09(1) & 761.87(1)  \\ \hline
{\tt SANC-2s}',~pid=24 & 1245.6(1) & 1239.6(1) & 763.11(1) & 761.88(1)  \\ \hline
\end{tabular}
\end{center}
\caption{\SANC comparison of decay width $\Gamma^{\rm Hard}$ in $\MeV$ in
the $\alpha(0)$ scheme for two options: $\Gamma_{t}=0(\neq 0)$; two variants:
5d and 2$s'$; and for two channels pid=22 and 24.
\label{table_tbud_hard_5_2_QCD}}
\vspace*{-3mm}
\end{table}

%------------------------------------
\subsubsection{One-loop decay width}
%------------------------------------

The numbers of this Section are produced with \SANC setup~(\ref{SANCinput}) in the
$\alpha(0)$ EW parameterization scheme. We present results for two processes: pid=22 and 24.

\vspace*{2mm}

$\bullet$ {pid = 22. Process $ t \rightarrow b + \mu^+ + \bar{\nu}_{\mu}$}

\vspace*{2mm}

In Table \ref{Table_tbud_corr_0l} we show the lowest-order width
 $\Gamma^{\rm Born}$ as a function of the $b$ quark mass.

\begin{table}[!h]
\begin{center}
\begin{tabular}{|l|l|l|l|}
\hline
$m_b,\GeV$             &  4.7         &   1.0         &   0.1         \\ \hline
$\Gamma^{\rm Born},\GeV$&  0.149094(1) &   0.149466(1) &   0.149483(1) \\ \hline
\end{tabular}
\end{center}
\vspace*{-5mm}
\caption{The total lowest-order width $\Gamma^{\rm Born}$ in $\GeV$ as function of $m_b$.
\label{Table_tbud_corr_0l}}
\vspace*{-3mm}
\end{table}

\noindent
As is seen, the effect of a finite $b$ quark mass is tiny. This justifies
considering the limit $m_b\to 0$, 
in which the formulas are much shorter and which is convenient to study the
validity of the KLN theorem in NLO approximation.\\

Table~\ref{Table11} demonstrates the stability of one-loop corrected quantities 
to the variation of the soft-hard separation parameter $E_{g}$.

\begin{table}[!h]
\begin{center}
\begin{tabular}{|l|l|l|l|}
\hline
\multicolumn{4}{|c|}{$\Gamma_{t}=0$}                         \\ \hline
$E_{g}$, GeV         &  $10^{-1}$  & $10^{-2}$   & $10^{-3}$  \\ \hline
$\Gamma^{\rm 1-loop}$ & 0.13651(1)  & 0.13652(1) & 0.13651(1) \\ \hline
$\delta\,\%$             & -8.440(1)   & -8.437(2)  & -8.439(9) \\ \hline   
\multicolumn{4}{|c|}{$\Gamma_{t}\neq0$}                       \\ \hline
$\Gamma^{\rm 1-loop}$ & 0.13735(1)  & 0.13734(1) & 0.13735(1) \\ \hline
$\delta \,\%      $ & -7.880(2)   & -7.883(2)  & -7.877(8)  \\ \hline
\end{tabular}
\end{center}
\caption[]{
The total one-loop corrected width $\Gamma^{\rm 1-loop}$ in GeV and relative one-loop correction 
$\delta=(\Gamma^{\rm 1-loop}-\Gamma^{\rm Born})/\Gamma^{\rm 1-loop}$ for
$E_{g} = 10^{-1},10^{-2},10^{-3}$~GeV and $m_b$=4.7 GeV.\label{Table11}}

\end{table}

As is seen from Table~\ref{Table11}, there is good $E_{g}$ stability for 
$E_{g} \leq 10^{-1}$~GeV.
However, there is a rather big difference  
of the results for the one-loop corrected width $\Gamma^{\rm 1-loop}$
and the relative QCD correction $\delta$ 
of the partial decay width between the calculation with the $\Gamma_{t}=0$
and the $\Gamma_{t}\neq0$ options, where in the latter case we use the
{\em fixed}$_{1}$ scheme defined in Ref.~\cite{Bardin:2010mz}.
 This difference gets smaller with formally decreasing $\Gamma_{t}$
as Table~\ref{Table_tbud_corr_1l} illustrates.

Table~\ref{Table_tbud_corr_1l} demonstrates the two dimensional convergence of 
$\Gamma^{\rm 1-loop}$ and $\delta$ when both $m_b$ and $\Gamma_{t}$ go to zero. 
Note that with decreasing $\Gamma_{t}$ the convergence in $m_b$ improves.

\begin{table}[!h]
\begin{center}
\begin{tabular}{|l|l|l|l|l|}
\hline
\multicolumn{5}{|c|}{$\Gamma^{\rm 1-loop},\GeV$}                \\ \hline
$\Gamma_{t}/10^N,\,N$&0&    1      &     2       &  $\Gamma_{t} = 0$
                                                               \\ \hline
$m_b,$ GeV &         &             &             &             \\ \hline
 4.7   & 0.137345(2) & 0.136594(2) & 0.136518(3) & 0.136515(3) \\ \hline 
 1.0   & 0.137919(2) & 0.136840(2) & 0.136726(3) & 0.136717(3) \\ \hline
 0.1   & 0.138479(2) & 0.136900(2) & 0.136738(3) & 0.136723(5) \\ \hline 
\multicolumn{5}{|c|}{$\delta,\% $}                             \\ \hline
$\Gamma_{t}/10^N,\,N$&0&     1     &     2       &  $\Gamma_{t} = 0$   
                                                               \\ \hline  
$m_b,$ GeV &         &             &             &             \\ \hline
 4.7   &  -7.881(1)  &  -8.384(1)  &  -8.436(2)  &  -8.437(2)  \\ \hline
 1.0   &  -7.726(1)  &  -8.448(1)  &  -8.523(2)  &  -8.530(2)  \\ \hline
 0.1   &  -7.362(1)  &  -8.418(1)  &  -8.526(2)  &  -8.537(3)  \\ \hline
\end{tabular}
\end{center}
\vspace*{-2mm}
\caption{
The total one-loop corrected width $\Gamma^{\rm 1-loop}$
in $\GeV$ and corresponding relative one-loop correction $\delta$ in 
$\%$ as function of the $b$ quark mass and of $\Gamma_{t}$
with soft-hard separation parameter $E_{g} = 10^{-2}\GeV$.}
\label{Table_tbud_corr_1l}
\end{table}

\newpage

$\bullet$ {pid = 24. Process $ t \rightarrow b + u + \bar{d}$}

\vspace*{2mm}

Here we present the corresponding set of Tables for the quark decay mode of
the top quark.

The total lowest-order width (Table \ref{Table_tbud_corr_0q}) also shows weak sensitivity 
to variations of the $b$ quark mass.

\begin{table}[!h]
\begin{center}
\begin{tabular}{|l|l|l|l|}
\hline
$m_b,\GeV$             &  4.7         &   1.0         &   0.1         \\ \hline
$\Gamma^{\rm Born},\GeV$&  0.447284(1) &   0.448399(1) &   0.448451(1) \\ \hline
\end{tabular}
\end{center}
\vspace*{-2mm}
\caption{The total lowest-order width $\Gamma^{\rm Born}$ in $\GeV$ as function of $m_b$.}
\label{Table_tbud_corr_0q}
\end{table}

The stability against variation of $E_{g}$ is shown in Table~\ref{Table12}.

\begin{table}[!h]
\begin{center}
\begin{tabular}{|l|l|l|l|}
\hline
\multicolumn{4}{|c|}{$\Gamma_{t}=0$}                      \\ \hline
$E_{g}$, GeV         & $10^{-1}$ & $10^{-2}$  & $10^{-3}$ \\ \hline
$\Gamma^{\rm 1-loop}$ & 0.3594(1) & 0.3581(1) & 0.3584(4) \\ \hline
$\delta$    \,\%     & -19.65(1) & -19.94(1) & -19.89(6)  \\ \hline
\multicolumn{4}{|c|}{$\Gamma_{t}\neq0$, {\em fixed}$_1$}  \\ \hline
$\Gamma^{\rm 1-loop}$ & 0.3619(1) & 0.3606(1) & 0.3607(3) \\ \hline
$\delta \,\%       $ & -19.10(1) & -19.37(1) & -19.35(7)  \\ \hline
\end{tabular}
\end{center}
\caption[]{
The total one-loop corrected width $\Gamma^{\rm 1-loop}$ in GeV and relative one-loop correction 
$\delta$ in $\%$ for $E_{g} = 10^{-1},10^{-2},10^{-3}$~GeV and $m_b$=4.7 GeV.\label{Table12}.}
\end{table}

Similar conclusions as drawn after Table~\ref{Table11} are valid in this case. Finally, 
Table~\ref{Table_tbud_corr_1q}, showing the two dimensional limit, demonstrates similar
behavior as Table~\ref{Table_tbud_corr_1l} for the semi-leptonic top decay.

\begin{table}[!h]
\begin{center}
\begin{tabular}{|l|l|l|l|l|}
\hline
\multicolumn{5}{|c|}{$\Gamma^{\rm 1-loop},\GeV$}               \\ \hline
$\Gamma_{t}/10^N,\,N$ &0&    1    &     2      &  $\Gamma_{t} = 0$   
                                                              \\ \hline  
$m_b,$ GeV &         &            &            &              \\ \hline  
 4.7   &  0.36054(8) & 0.35830(8) & 0.35808(8) & 0.35812(6)   \\ \hline
 1.0   &  0.36210(8) & 0.35890(8) & 0.35863(8) & 0.35870(6)   \\ \hline
 0.1   &  0.36380(8) & 0.35898(8) & 0.35854(8) & 0.35861(7)   \\ \hline
\multicolumn{5}{|c|}{$\delta,\% $} \\ \hline
$\Gamma_{t}/10^N,\,N$ &0&    1    &     2      &  $\Gamma_{t} = 0$   
                                                              \\ \hline  
$m_b,$ GeV &         &            &            &              \\ \hline
 4.7   &  -19.39(2)  & -19.89(2)  & -19.94(2)  & -19.94(1)    \\ \hline
 1.0   &  -19.25(2)  & -19.96(2)  & -20.02(2)  & -20.00(1)    \\ \hline
 0.1   &  -18.88(2)  & -19.95(2)  & -20.05(2)  & -20.03(2)    \\ \hline
\end{tabular}
\end{center}
\caption{
The total one-loop corrected width $\Gamma^{\rm 1-loop}$
in $\GeV$ and corresponding relative one-loop correction $\delta$ in 
$\%$ for $E_{g} = 10^{-2}\GeV$ as function of the $b$ quark mass and of $\Gamma_{t}$.}
\label{Table_tbud_corr_1q}
\end{table}

\newpage

\subsection{$s$ channel}
%-----------------------
Here we consider $s$ channel processes:

\vspace*{2mm}

$\bullet$ pid=25, $\bar{d} +u \rightarrow \bar{b}  + t$
and pid=26, $\bar{u} +d \rightarrow \bar{t}  + b$ 

\vspace*{2mm}

but it is sufficient to consider one channel, say pid=25.

\subsubsection{SANC--CompHEP comparison}
%---------------------------------------

\begin{table}[!h]
\begin{center}
\begin{tabular}{|l|l|l|l|l|l|}
\hline
$\sqrt{\hs},\GeV$ &  200      &  1000       &  7000           \\ \hline
\multicolumn{4}{|c|}{$\Gamma_{t}=0$}                          \\ \hline
{\tt CompHEP}  &  0.72930(2) &   1.6340(1) &    0.071223(1)  \\ \hline
{\tt SANC-4d}  &  0.72927(1) &   1.6341(1) &    0.071229(1)  \\ \hline
{\tt SANC-s}'  &  0.72928(1) &   1.6338(1) &    0.071227(1)  \\ \hline
\multicolumn{4}{|c|}{$\Gamma_{t}\neq0$}                       \\ \hline
{\tt CompHEP}  &  0.72638(2) &   1.6323(1) &    0.070946(1)  \\ \hline 
{\tt SANC-4d}  &  0.72636(1) &   1.6322(1) &    0.070942(1)  \\ \hline
{\tt SANC-s}'  &  0.72636(1) &   1.6322(1) &    0.070942(1)  \\ \hline
\end{tabular}                                              
\end{center}
\caption{
Comparison of the cross section $\sigma^{\rm Hard}(\sqrt{\hs},~\Gamma_{t})$ in pb
for three cms energies and two width options  $\Gamma_{t}=0 (\neq 0)$.
for pid = 25, i.e. for process~ $\bar{d} +u \rightarrow \bar{b}  + t$;
the $E_g$ cut is equal to 2 GeV.
\label{table_udtb_hard_QCD}}
\end{table}

As is seen from Table~\ref{table_udtb_hard_QCD}, there is very good agreement 
of numbers obtained from \SANC and {\tt CompHEP} within the statistical
errors for two considered width options and three cms energies.
In this case, the two widths options agree well for all three energies.

%-----------------------------------
\subsubsection{One-loop corrections}
%-----------------------------------

The analogue of Tables 4--9 of~Ref.~\cite{Bardin:2010mz} is now represented by one joint 
Table~\ref{table_udtb_corr_QCD_2} showing the stability of one-loop corrected
QCD cross sections $\hat{\sigma}^{\rm 1-loop}$ and relative QCD RC $\delta$ against variation
of the soft-hard separation
parameter, $\bar{\omega}$, and the difference between the two options: $\Gamma_{t}=0(\neq0)$.
\begin{table}[!h]
\begin{center}
\begin{tabular}{|l|l|l|l|l|l|l|}
\hline
$\sqrt{s}, \GeV$      & \multicolumn{2}{|c|}{200}
                      & \multicolumn{2}{|c|}{1000}
                      & \multicolumn{2}{|c|}{7000} \\ \hline
$\sigma^{\rm Born}$,fb & \multicolumn{2}{|c|}{308.09055(1)}
                      & \multicolumn{2}{|c|}{105.97718(1)}
                      & \multicolumn{2}{|c|}{2.2352950(1)}
\\ \hline
$ \bar \omega$          & $10^{-5}$ & $10^{-6}$ & $10^{-5}$ & $10^{-6}$ & $10^{-5}$ & $10^{-6}$
\\ \hline
\multicolumn{7}{|c|}{$\wtp=0, \mq$}			  			
\\ \hline
$\hat\sigma^{\rm 1-loop}$&-374.05(1) &-374.07(1)& 0.36446(1)& 364.46(1) &17.599(1) & 17.599(1)
\\ \hline
$\delta,\%$             & -221.41(1)& -221.42(1)& 243.90(1) & 243.90(1) &687.34(1) & 687.34(1) 
\\ \hline
$\hat{\sigma}^{\MSbar}$  & 525.34(1) & 525.37(1)& 0.17057(1)& 170.57(1) &8.2097(1) & 8.2097(1)
\\ \hline 
$\delta^{\MSbar},\%$     & 70.51(1)  & 70.53(1) & 60.95(1)  & 60.95(1)  &267.28(1) & 267.28(1)  
\\ \hline
\multicolumn{7}{|c|}{$\wtp=0, \mq/10$}
\\ \hline
$\hat{\sigma}^{\MSbar}$  & 525.34(1) & 525.37(1)& 0.17057(1)& 170.57(1) &8.2097(1) & 8.2097(1) 
\\ \hline
$\delta^{\MSbar},\%$     &  70.51(1) & 70.52(1) & 60.95(1)  & 60.95(1)  &267.27(1) & 267.28(1)
\\ \hline
\multicolumn{7}{|c|}{$\wtp \neq 0, \mq $}
\\ \hline 
$\sigma^{\rm 1-loop}$    &-372.85(1)&-372.87(1) & 0.36455(1)& 364.55(1) &17.601(1) & 17.601(1) 
\\ \hline 
$\delta,~\% $           &-221.02(1) &-221.03(1) & 243.99(1) & 243.99(1) &687.40(1) & 687.40(1) 
\\ \hline
$\hat{\sigma}^{\MSbar}$  & 526.54(1)& 526.57(1) &0.17066(1) & 170.66(1) &8.2111(1) & 8.2111(1)
\\ \hline
$\delta^{\MSbar},\%$     & 70.90(1) &  70.91(1) & 61.03(1)  & 61.03(1)  &267.34(1) & 267.34(1)
\\ \hline
\multicolumn{7}{|c|}{$\wtp \neq 0, \mq/10$}	
\\ \hline
$\hat{\sigma}^{\MSbar}$  & 526.54(1)& 526.57(1) &0.17066(1) & 170.66(1) &8.2111(1) & 8.2111(1)
\\ \hline
$\delta^{\MSbar},\%$     & 70.90(1) &  70.91(1) & 61.03(1)  & 61.03(1)  &267.34(1) & 267.34(1)
\\ \hline
\end{tabular}
\end{center}
\caption{The total lowest-order cross section $\sigma^{\rm Born}$,
one-loop cross section $\hat\sigma^{\rm 1-loop}$ and total one-loop corrected $\overline{MS}$ 
subtracted quantities $\hat{\sigma}^{\MSbar}$ and corresponding $\delta$ and $\delta^{\MSbar}$ 
in $\%$ at  $\omega=10^{-5},10^{-6}$ for $\sqrt{s} = 200, 1000, 7000 \GeV$.
\label{table_udtb_corr_QCD_2}}
\end{table}

As is seen, there is good $\bar{\omega}$ stability in the interval
 $\bar{\omega}=10^{-5}\div 10^{-6}$; also, the subtracted
quantities $\hat{\sigma}^{\MSbar}$ and $\delta^{\MSbar}$ for all
three cms energies and for both width options, $\Gamma_{t}=0(\neq0)$,
are seen to be independent of the light quark masses.

One can see that the difference between the two width options is of order of few 
per mille in absolute deviation, rapidly decreasing with increasing cms energy.
We therefore conclude that one may use the usual infrared regularization for 
$s$ channel processes as was the case for EW corrections, see text after Table 6 of 
Ref.~\cite{Bardin:2010mz}.

In Table~\ref{Table_udtb_corr_1sub} we demonstrate two dimensional convergence of 
$\hat{\sigma}^{\MSbar}_1$ and $\delta^{\MSbar}$ when both $m_b$ and $\Gamma_{t}$
go to zero. Note that, similarly to top decay cases, 
with lowering of $\Gamma_{t}$ the convergence in $m_b$ improves.

The agreement of $\hat{\sigma}^{\MSbar}_1$ and $\delta^{\MSbar}$
in the double limit (numbers for $\Gamma_{t}/10^4$ and $\Gamma_{t} = 0$) illustrates our
analytic considerations described in Section~\ref{BVudtb}.

\begin{table}[!h]
\begin{center}
\begin{tabular}{|l|l|l|l|l|}
\hline
\multicolumn{5}{|c|}{$\hat{\sigma}^{\MSbar}_1$, pb}            \\ \hline
$\Gamma_{t}/10^N,\,N$ &0&     2      &     4      &  $\Gamma_{t} = 0$   \\ \hline  
$m_b,$ GeV    &         &            &            &            \\ \hline  
 4.7   &     0.52654(1) & 0.52530(1) & 0.52534(1) & 0.52534(1) \\ \hline  
 1.0   &     0.53099(1) & 0.52783(1) & 0.52776(1) & 0.52778(1) \\ \hline  
 0.1   &     0.53386(1) & 0.52791(1) & 0.52784(1) & 0.52784(1) \\ \hline  
 0.01  &     0.53656(1) & 0.52793(1) & 0.52783(1) & 0.52785(1) \\ \hline  
\multicolumn{5}{|c|}{$\delta^{\MSbar},\% $}                     \\ \hline
$\Gamma_{t}/10^N,\,N$ &0&     2      &     4      &  $\Gamma_{t} = 0$   \\ \hline  
$m_b,$ GeV    &         &            &            &            \\ \hline
 4.7   &     70.90(1)   & 70.50(1)   & 70.52(1)   &   70.51(1) \\ \hline
 1.0   &     69.65(1)   & 68.64(1)   & 68.62(1)   &   68.62(1) \\ \hline
 0.1   &     70.44(1)   & 68.54(1)   & 68.52(1)   &   68.52(1) \\ \hline
 0.01  &     71.30(1)   & 68.55(1)   & 68.52(1)   &   68.52(1) \\ \hline
\end{tabular}
\end{center}
\caption{
The total one-loop corrected $\MSbar$ subtracted cross sections $\hat{\sigma}^{\MSbar}$ in pb 
and corresponding relative one-loop correction $\delta^{\MSbar}$ in $\%$ for $\sqrt{\hs}$=200$\GeV$
(at the parton level) 
and for $\bar{\omega} = 10^{-5}$ as function of the $b$ quark mass and of $\Gamma_{t}$.
\label{Table_udtb_corr_1sub}}
\end{table}
%----------

\clearpage

\subsection{$t$ channel}
%-----------------------

\subsubsection{SANC--CompHEP comparison}
%---------------------------------------

This is the most complicated case: $t$ channel cross sections show up bad
statistical convergence. For this comparison we use CompHEP v.4.5.1 and its setup,
but with non-zero masses of the $u$ and $d$ quarks
(accessed via $bc\to ts\gamma$ channel).
For the Tables of this subsection we used $m_q=m_u=m_d= 66$ MeV
and $10m_q$, i.e. $m_u=m_d= 660$ MeV. 
The cut on the cms gluon energy was $E_g\geq$ 2 GeV, and $\alpha_s = 0.12201\,(Q=\mz)$.
As in~\cite{Bardin:2010mz}, rows marked ``{\tt SANC(S)}'' (``{\tt S}''hort massive case)
were computed retaining $m_q$ or $10m_q$ only in fermion propagators radiating
a gluon, while ``{\tt SANC(F)}'' (``{\tt F}''ully massive case) means that light quark
masses were kept everywhere. 

\vspace*{2mm}

$\bullet$ {pid = 27: Process $ b + u  \to d + t$}

\begin{table}[!h]
\begin{center}
\begin{tabular}{|l|c|c|c|}
\hline
$\sqrt{\hs}/$GeV& 200   & 1000     & 7000      \\ \hline
\multicolumn{4}{|c|}{$\Gamma_{t}=0$, $m_q$}       \\ \hline
{\tt CompHEP}&   17.908(1)  &  415.04(6)  &  617.40(6)  \\ \hline
{\tt SANC(S)}&   17.904(1)  &  416.39(1)  &  618.94(1)  \\ \hline
\multicolumn{4}{|c|}{$\Gamma_{t}\neq 0$, $m_q$}   \\ \hline
{\tt CompHEP}&   17.761(1)  &  404.29(6)  &  574.02(6)  \\ \hline
{\tt SANC(S)}&   17.759(1)  &  405.62(1)  &  575.61(1)  \\ \hline      
{\tt SANC(F)}&   17.760(1)  &  405.63(1)  &  575.62(1)  \\ \hline
\multicolumn{4}{|c|}{$\Gamma_{t}\neq 0$, $10m_q$} \\ \hline
{\tt CompHEP}&   11.792(1)  &  286.79(7)  &  400.29(2)  \\ \hline
{\tt SANC(F)}&   11.789(1)  &  287.08(1)  &  400.48(1)  \\ \hline
\end{tabular}
\end{center}
\caption[]{Comparison of the cross section $\hat{\sigma}^{\rm{hard}}(\sqrt{\hs})$ in pb
for the process $b+u\to t+d$ for three cms energies;
three options of four: $(\Gamma_{t}=0(\neq0))\otimes(m_q,10m_q)$.
\label{Table31}}
\end{table}
%----------

$\bullet$ {pid = 28: Process $ b + \bar{d} \to \bar{u} + t$}

\begin{table}[!h]
\begin{center}
\begin{tabular}{|l|c|c|c|}
\hline
$\sqrt{\hs}/$GeV& 200   & 1000     & 7000      \\ \hline
\multicolumn{4}{|c|}{$\Gamma_{t}=0$, $m_q$}       \\ \hline
{\tt CompHEP}&   10.335(1)  &  388.74(7)  &  615.5(1)   \\ \hline
{\tt SANC(S)}&   10.332(1)  &  389.87(1)  &  616.86(1)  \\ \hline
\multicolumn{4}{|c|}{$\Gamma_{t}\neq 0$, $m_q$}   \\ \hline
{\tt CompHEP}&   10.245(1)  &  378.21(7)  &  572.0(1)   \\ \hline
{\tt SANC(S)}&   10.241(1)  &  379.34(1)  &  573.58(1)  \\ \hline
{\tt SANC(F)}&   10.241(1)  &  379.35(1)  &  573.59(1)  \\ \hline
\multicolumn{4}{|c|}{$\Gamma_{t}\neq 0$, $10m_q$} \\ \hline
{\tt CompHEP}&   6.641(1)   &  266.75(2)  &  398.68(4)  \\ \hline
{\tt SANC(F)}&   6.640(1)   &  266.83(1)  &  398.83(1)  \\ \hline
\end{tabular}
\end{center}
\caption[]{The same as Table~\ref{Table31}, but for the process
$ b + \bar{d} \to \bar{u} + t$.
\label{Table32}}
\end{table}
%----------

The results of Tables~\ref{Table31}--\ref{Table32} are qualitatively the same
and may be discussed together.

There is satisfactory agreement for all options only near the reaction threshold. At higher
energies {\tt SANC/CompHEP} agree poorly for the $m_q$ option and much better for the $10m_q$ 
option.
This allows us to draw conclusions similar to those which were drawn at the discussion 
of the results of analogous Tables 10--11 of Ref.~\cite{Bardin:2010mz}.

\subsubsection{One-loop QCD corrections}
%--------------------------------------

The numerical results for this subsection were again produced with 
the aim to demonstrate the stability of one-loop corrected QCD cross sections
$\sigma^{\rm 1-loop}$ and relative QCD RC $\delta$ against variation of the soft-hard separator
$\bar{\omega}$ and to study the difference between the two options: $\Gamma_{t}=0(\neq0)$.

\vspace*{2mm}

$\bullet$ {pid = 27. Process $ b + u  \to d + t$}

\vspace*{2mm}

Table~\ref{table_budt_corr_QCD_2} is the counterpart for QCD of the six Tables 12--17 of 
Ref.~\cite{Bardin:2010mz}.

\begin{table}[!h]
\begin{center}
\begin{tabular}{|l|l|l|l|l|l|l|}
\hline
$\sqrt{s}, \GeV$        & \multicolumn{2}{|c|}{200} 
                        & \multicolumn{2}{|c|}{1000} 
                        & \multicolumn{2}{|c|}{7000} \\ \hline
$\sigma^{\rm Born},$ pb  & \multicolumn{2}{|c|}{7.3551153(1)}   
                        & \multicolumn{2}{|c|}{48.993407(2)}
                        & \multicolumn{2}{|c|}{50.824228(2)} 
\\ \hline
$ \bar \omega$          & $10^{-5}$  &  $10^{-6}$ & $10^{-5}$ &  $10^{-6}$ &  $10^{-5}$ & $10^{-6}$
\\ \hline
\multicolumn{7}{|c|}{$\wtp=0, \mq$}	
\\ \hline 
$\hat\sigma^{\rm 1-loop}$& -5.741(1)  & -5.742(1)  & 43.67(1)  &  43.66(1) &  50.57(2)  &  50.57(2)
\\ \hline
$\delta,\%$             & -178.06(1) & -178.07(1) & -10.90(2) & -10.88(3) &   -0.50(3) &  -0.48(3)
\\ \hline
$\hat{\sigma}^{\MSbar}$  & 10.215(1)  & 10.215(1)  &  47.67(1) & 47.67(1)  &  50.71(2)  &  50.72(2)
\\ \hline 
$\delta^{\MSbar},\%$     & 38.88(1)   & 38.88(1)   &  -2.70(2) &  -2.69(3) &   -0.23(3) &   -0.20(4)
\\ \hline				    			  			
\multicolumn{7}{|c|}{$\wtp=0, \mq/10$}
\\ \hline 
$\hat{\sigma}^{\MSbar}$  & 10.212(1) & 10.210(1)   &   47.66(2)&  47.66(2)  &  50.71(2) &   50.72(3)
\\ \hline
$\delta^{\MSbar},\%$     & 38.88(2)  & 38.86(2)    & -2.73(4)  &   -2.72(4) & -0.22(5)  &   -0.21(6)
\\ \hline				    			  			
\multicolumn{7}{|c|}{$\wtp \neq 0, \mq $}	
\\ \hline 
$\sigma^{\rm 1-loop}$    & -5.573(1)  & -5.574(1)  & 43.94(1)   & 43.95(1)   & 50.87(2)  &  50.86(2)
\\ \hline 
$\delta,~\% $           & -175.77(1) & -175.78(1) & -10.29(3)  & -10.28(2)  & 0.097(29) & 0.063(34)
\\ \hline
\end{tabular}
\end{center}
\caption{The total lowest-order cross section $\sigma^{\rm Born}$, one-loop  cross section 
$\hat\sigma^{\rm 1-loop}$ and total one-loop corrected $\overline{MS}$ subtracted 
quantities $\hat{\sigma}^{\MSbar}$ (all in pb) and corresponding $\delta$ and $\delta^{\MSbar}$ 
in $\%$ for $\bar{\omega}=10^{-5},10^{-6}$ and $\sqrt{s} = 200, 1000, 7000 \GeV$.}
\label{table_budt_corr_QCD_2}
\end{table}

Stability in $\bar{\omega}$ is seen to hold in all considered variants;
$\mq$ independence is clearly seen for the $\wtp=0$ option. However, contrary to EW corrections,
the dependence on $\wtp$ itself for non-subtracted quantities (compare two $\delta$ rows) 
is much more pronounced, varying from several $\%$ near the threshold to half a percent at higher 
energies. Since it is not known how to realize subtraction for $\wtp\ne 0$ option, this difference
may be treated as a theoretical uncertainty of NLO calculations which is,
however, much smaller than the estimate of NNLO contributions~\cite{Kidonakis:2011tg}.

For $t$ channel processes, we show only one-dimensional convergence in the limit $\wtp\to 0$ for 
$\sqrt{s} = 200 \GeV$ where our NLO correction is unphysical ($< -100\%$), see 
Table~\ref{table_budt_corr_QCD_1}. (We recall, that in this case we do not have
an analytic proof of convergence similar to the proof
shown in Section~\ref{BVudtb} for the $s$ channel).
So, it should be considered as only a formal, numerical illustration of convergence. 

\begin{table}[!h]
\begin{center}
\begin{tabular}{|l|l|l|} 
\hline
\multicolumn{3}{|c|}{  
 $\sqrt{s} = 200 \GeV , \sigma^{\rm Born} = 7.3551153(1)$ pb}
\\ \hline
$\wtp/10^N,N$ &  $\hat\sigma^{\rm 1-loop}$, pb  &   $\delta$,\%
\\ \hline
 0            & -5.5730(6) & -175.77(1)
\\ \hline
 2            & -5.7392(6) & -178.03(1)
\\ \hline
 4            & -5.7419(7) & -178.07(1)
\\ \hline \hline
$\wtp = 0$    & -5.7411(6) & -178.06(1)
\\ \hline
\end{tabular}
\end{center}
\caption{The total lowest-order cross section $\sigma^{\rm Born}$,
the total one-loop corrected $\hat\sigma^{\rm 1-loop}$ in pb
and corresponding $\delta$ in $\%$ at $\sqrt{s} = 200 \GeV$ 
and $\wtp/10^N, N=0,2,4.$}
\label{table_budt_corr_QCD_1}
\end{table}

\vspace*{2mm}

$\bullet$ {pid = 28. Process $ b + \bar{d} \to \bar{u} + t$}

\vspace*{2mm}

Table~\ref{table_budt_corr_QCD_4} is the counterpart for QCD of the
six Tables 18--22 of Ref.~\cite{Bardin:2010mz}. 

\begin{table}[!h]
\begin{center}
\begin{tabular}{|l|l|l|l|l|l|l|}
\hline
$\sqrt{s},~\GeV$       & \multicolumn{2}{|c|}{200 }   
                       & \multicolumn{2}{|c|}{1000}
                       & \multicolumn{2}{|c|}{7000} \\ \hline  
$\sigma^{\rm Born},$ pb & \multicolumn{2}{|c|}{4.49579065(2)}   
                       & \multicolumn{2}{|c|}{46.6869560(1)}
                       & \multicolumn{2}{|c|}{50.72449(1)}  

\\ \hline 
$  \bar\omega$         & $10^{-5}$  &  $10^{-6}$ & $10^{-5}$ &  $10^{-6}$ &  $10^{-5}$  & $10^{-6}$
\\ \hline
\multicolumn{7}{|c|}{$\wtp = 0, \mq $}
\\ \hline
$\hat\sigma^{\rm 1-loop}$&-3.2657(4) & -3.2666(6)&   39.71(1) & 39.70(1)  &  50.22(2) &  50.22(2)
\\ \hline
$\delta,\%$             & -172.64(1)& -172.66(1)&  -14.95(2) & -14.96(3) &  -0.99(3) &  -1.00(4)
\\ \hline
$\hat{\sigma}^{\MSbar}$  & 6.5587(4) & 6.5586(6) &  45.29(1)  & 45.29(1)  &  50.53(2) &  50.52(2)
\\ \hline 
$\delta^{\MSbar},\%$     &  45.88(1) &  45.88(1) &  -2.98(2)  & -2.99(3)  &  -0.38(3) &  -0.39(4)
\\ \hline	
\multicolumn{7}{|c|}{$\wtp = 0, \mq/10 $}
\\ \hline 
$\hat{\sigma}^{\MSbar}$  & 6.5564(7) & 6.5553(9) & 45.28(2)   & 45.28(2)   & 50.51(3) & 50.53(3)
\\ \hline 
$\delta^{\MSbar},\%$     & 45.91(2)  & 45.88(2)  &  -3.01(4)  & -3.02(4)   & -0.42(5) & -0.39(6)
\\ \hline				    			  			
\multicolumn{7}{|c|}{$\wtp \neq 0, \mq $}
\\ \hline 
$\hat\sigma^{\rm 1-loop}$& -3.1526(4) & -3.1532(5)&  39.99(1)  &  39.99(1) &  50.52(1) & 50.51(2)
\\ \hline
$\delta,~\%$            & -170.12(1) &-170.14(1) &  -14.35(2) & -14.34(3) & -0.40(3)  & -0.41(3)
\\ \hline
\end{tabular}
\end{center}
\caption{The same as Table~\ref{table_budt_corr_QCD_2} but for 
pid = 28. Process $ b + \bar{d} \to \bar{u} + t$.\label{table_budt_corr_QCD_4}}
\end{table}

\begin{table}[!h]
\begin{center}
\begin{tabular}{|l|l|l|}
\hline
\multicolumn{3}{|c|}{ $\sqrt{s} = 200 \GeV , \sigma^{\rm Born} = 4.495790646(3)$ pb}   
\\ \hline
$\wtp/10^N,N$  & $\sigma^{1-loop}$  &  $\delta$
\\ \hline
0              & -3.1526(4) & -170.12(1)
\\ \hline 
2              & -3.2644(4) & -172.61(1)
\\ \hline
4              & -3.2661(5) & -172.65(1)
\\ \hline
$\wtp=0$       & -3.2657(4) & -172.64(1)
\\ \hline
\end{tabular}
\end{center}
\caption{The same as Table~\ref{table_budt_corr_QCD_1} but for 
pid = 28. Process $ b + \bar{d} \to \bar{u} + t$.\label{table_budt_corr_QCD_3}}
\end{table}

From Tables~\ref{table_budt_corr_QCD_4}--\ref{table_budt_corr_QCD_3} the same conclusions may be
drawn as from Tables~\ref{table_budt_corr_QCD_2}--\ref{table_budt_corr_QCD_1}. We will not
repeat them here.

\clearpage

\section{Conclusions and Outlook\label{concl}}
%--------------------------------------------- 
In two papers, this one and Ref.~\cite{Bardin:2010mz}, we have described the implementation into 
the \SANC framework of the complete one-loop QCD and EW calculations, including hard bremsstrahlung
contributions, for the processes of top quark decays and of $s$ and $t$ channel production, 
the latter two at the partonic level.
Within the \SANC framework, we have created the standard FORM and FORTRAN modules,
see,~Ref.\cite{Andonov:2008ga}, compiled into a package {\tt sanc\_cc\_v1.40} which
may be downloaded from the \SANC project homepages~\cite{homepagesSANC}.
 
The essentially new aspect of these two papers is the study of regularisation of the top 
quark--photon and --gluon infrared divergences with aid of the complex mass of the top quark. 
A comparison of these NLO corrections computed within this approach 
with those computed by the conventional method showed a difference which is not negligible, 
though not exceeding $\sim 1\%$ for both types of corrections and for three considered processes.
Although, the introduction of the non-zero top width within the fixed width scheme violates
gauge-invariance, it is numerically not so important being of the order ${\cal{O}}(\wtp/\mtp)$;
it restores in the limit $\wtp\to 0$.

The emphasis of these papers is to be assured of the correctness of our results. 
We observe the independence of the form factors on gauge parameters, checked the stability 
of the result against variation of the soft-hard separation parameter $\bar{\omega}$ and 
the independence of the $\MSbar$ subtracted quantities off the initial quark
masses which is crucial for calculations at the hadronic level.

As has become \SANC standard, we have compared our numerical results with other independent 
calculations wherever possible. For the decay channels it was done  in our previous papers 
(both EW~\cite{Arbuzov:2007ke} and QCD~\cite{Andonov:2007zz}), showing good agreement.
As usual, the Born level and the hard gluon contributions of all three
channels were checked against the {\tt CompHEP} package. We found very
good agreement for both $\Gamma_{t}=0(\neq0)$ options.

As far as the comparison of EWRC for the production processes is concerned,
it was described in our previous paper Ref.~\cite{Bardin:2010mz} and we found no
suitable results in the literature on QCD corrections at the partonic level.

As a first step toward comparison of QCD calculations at the hadronic level we performed
a triple comparison of the LO total cross sections of $s$ and $t$ channel single top production 
computed by {\tt SANC/CompHEP/MCFM} with the same set of PDF. We found a satisfactory agreement
between {\tt SANC} and {\tt CompHEP} numbers. The comprehensive comparison of the results
at the hadronic level is the subject of ongoing work and a further publication.

All the calculations were done using a combination of analytic and Monte
Carlo integration methods which will make it easy to impose experimental
cuts in future calculations for  $pp$ collisions at the hadronic level.
The results presented in these two papers lay a solid base for future
extensions of calculations for the single top production channels at the LHC 
with subsequent decay in the cascade (factorized) approximation (see e.g.~\cite{Bardin:2009wv})
which simultaneously take account of NLO EW and QCD corrections.

\bigskip\noindent
{\bf Acknowledgements.}
This work is partly supported by Russian Foundation for Basic Research
grant $N^{o}$ 10-02-01030. 

WvS is indebted to the directorate of the Dzhelepov Laboratory of Nuclear
Problems, JINR, Dubna, for the hospitality extended to him during April 2011.

\clearpage

\bibliographystyle{utphys_spires}
\addcontentsline{toc}{section}{\refname}\bibliography{Top_EW}
\end{document}